\documentclass[useAMS,usenatbib]{mn2e}

\usepackage{natbib}
\bibpunct{(}{)}{;}{a}{,}{,}

\usepackage{graphicx}
\usepackage{helvet}
\usepackage{natbib}
\usepackage{setspace}
\usepackage{amssymb}
\usepackage{longtable}

\title[Main-sequence variable stars]
{Main-sequence variable stars in young open cluster NGC 1893}

\author[Sneh Lata et al.]
       {Sneh Lata$^1$\thanks{E-mail: sneh@aries.res.in}, Ram Kesh Yadav$^1$, A. K. Pandey$^1$,  Andrea Richichi$^2$,  C. Eswaraiah$^{1, 3}$,  
\newauthor Brajesh Kumar$^{1, 4}$, Norbert Kappelmann$^5$ and Saurabh Sharma$^1$ \\ 
       $^1$Aryabhatta Research Institute of Observational Sciences, Manora Peak, Nainital 263002, Uttarakhand, India \\
       $^2$National Astronomical Research Institute of Thailand, 191 Siriphanich Bldg., Huay Kaew Rd., Suthep, Muang, Chiang Mai
50200, Thailand \\
       $^3$Institute of Astronomy, National Central University, 300 Jhongda Rd, Jhongli, Taoyuan Country 32054, Taiwan \\
        $^{4}$Institut d'Astrophysique et de G\'{e}ophysique, Universit\'{e} de Li\`{e}ge, All\'{e}e du 6 Ao\^{u}t 17, B\^{a}t B5c,
      4000 Li\`{e}ge, Belgium \\
       $^5$Institute of Astronomy and Astrophysics, Sand 1, 72076, T$\ddot{u}$bingen, Germany \\}

\date{Accepted ---------.
      Received ---------;
      }

\pagerange{\pageref{firstpage}--\pageref{lastpage}}

\def\LaTeX{L\kern-.36em\raise.3ex\hbox{a}\kern-.15em
    T\kern-.1667em\lower.7ex\hbox{E}\kern-.125emX}

\begin{document}

\label{firstpage}

\maketitle

\label{firstpage}
\begin{abstract}
In this paper 
we present 
time series photometry of 104 variable stars in the cluster
region NGC 1893. The association of the present variable candidates to the cluster NGC 1893 has been determined by using $(U-B)/(B-V)$ and $(J-H)/(H-K)$ two colour diagrams, and
$V/(V-I)$ colour magnitude diagram.
Forty five stars are found to be main-sequence variables and these could
be B-type variable stars associated with the cluster. 
We classified these objects as $\beta$ Cep, slowly pulsating B stars and new class variables as discussed by Mowlavi et al. (2013).
These variable candidates show $\sim$0.005 to $\sim$0.02 mag brightness variations with periods of $<$ 1.0 d.
Seventeen new class variables are located in the
$H-R$ diagram between the slowly pulsating B stars and 
$\delta$ Scuti variables.
Pulsation could be one of the causes for periodic brightness variations in these stars.
The X-ray emission of present main-sequence variables associated with the cluster lies in the saturated region of X-ray luminosity versus period diagram and follows
the general trend by Pizzolato et al. (2003).
\end{abstract}

\begin {keywords} 
Open  cluster:  NGC 1893  --
colour--magnitude diagram: Variables-main sequence stars-B type 
\end {keywords}

\section{Introduction}
The aim of the present study is to analyze the light curves of those stars which lie in the upper part of the main-sequence (MS) in the colour-magnitude diagram (CMD) of NGC 1893. 
The upper part of the MS consists of stars of spectral type O to A. 
The brightness variation in OB supergiants, early B-type stars, Be stars, mid to late B-type stars 
occurs mostly due to the pulsations (Stankov \& Handler 2005; Kiriakidis et al. 1992; Moskalik \& Dziembowski 1992).
Pulsating variable stars expand and contract in a repeating cycle of size changes. 
The different types of pulsating variables are distinguished by their periods of pulsation and the shapes of their light curves. 
These could be $\beta$ Cep, slowly pulsating B (SPB), $\delta$ Scuti stars etc.
The $\beta$ Cep stars are pulsating MS variables and found to be lying above the upper MS in the $H-R$ diagram with early B spectral types (Handler \& Meingast 2011).
These have periods and amplitude in the range of 0.1 - 0.6 d and 0.01 to 0.3 mag, respectively.

The SPB stars lie 
just below the instability strip of $\beta$ Cep variables (Waelkens 1991). 
The SPB stars are nearly perfectly confined to the main-sequence band (e.g., De Cat et al. 2004). 
The SPB stars are less massive (3-7 $M_{\odot}$) in comparison to $\beta$ Cep stars (8-18 $M_{\odot}$).
The effective temperature of known SPB stars lies in the range of 10000 to 20000  K. The well known Kappa mechanism (Dziembowski et al. 1993; Gautschy \& Saio 1993) is the reason for the periodic brightness variation in these stars.
The SPB stars are slow pulsator with period of more than 0.5 d. 
Theoretical instability strip of SPB stars overlap with instability strip of $\beta$ Cep stars.
Waelkens et al. (1998) classified a huge number of B-type stars  as new SPB stars using the Hipparcos mission.

Another class of stars populating with the B-type
MS are the Be stars. They are defined as non-
supergiant B star with one or more Balmer lines in their emission.
Classical Be stars are physically known as
rapidly rotating B-type stars with line emission.
As pulsating Be stars occupy the same region of the $H-R$ diagram as $\beta$ Cep and SPB stars, it is generally assumed
that pulsations in Be stars have the same origin as the case of $\beta$ Cep and SPB stars (Diago et al. 2008).

The another group of pulsation variables is $\gamma$ Dordus (period: 0.3-1.0 d). They are found to be located below the instability strip of $\delta$ Scuti stars. The instability strip of $\gamma$ Dordus
overlaps with instability strip of $\delta$ Scuti. The $\delta$ Scuti stars (these are short period variables with periods lying in the range of 0.03 to 0.3 d) 
  are part of the classical instability strip where
Cepheids are found and these Cepheids
are radially pulsating, high luminosity (classes Ib-II) variables with periods in the range of
1-135 days and amplitudes from several hundredths to $\sim $2 mag in $V$ (the amplitudes are
greater in the $B$ band). The spectral type of these objects at maximum light is F whereas at the minimum, the spectral types are G-K. The longer the period
of light variation, the later is the spectral type. 

In addition to above discussions we would like to give a brief description of previous studies on B-type stars' 
variability. 
The study of pulsating B-stars having same age and chemical composition
in young open clusters provides understanding to interpret variability (Handler et al. 2008; Majewska et al. 2008;
Michalska et al. 2009; Handler \& Meingast 2011; Jerzykiewicz et al. 2011; Saesen
 et al. 2013; Balona
et al. 1997; Gruber et al. 2012).
Saesen et al. (2013) using the photometric study of the B-type stars in NGC 884 combined with their recent spectroscopic observations
 offer an interesting and different approach
to the advancement of understanding of these young massive objects.
Saesen et al. (2010) presented differential time-resolved multi-colour CCD photometry of NGC 884 cluster that leads to 
the identification of 36 multi-periodic and 39 mono-periodic B-stars, 19
multi-periodic and 24 mono-periodic A- and F-stars, and 20 multi-periodic and 20 mono-periodic variable
stars of unknown nature. 
Saio et al. (2006) used MOST (Microvariability and Oscillations
of Stars) satellite to detect variability in supergiant star HD 163899 (B2 Ib/II), and they found 48 frequencies with amplitudes of a few millimagnitudes (mmag) and less. The frequency range is similar to g- and p-mode pulsations.

Balona et al. (2011) presented Kepler observations of variability in B-type stars. They presented the light curves of 48 B-type stars.
They find no evidence for pulsating stars between the cool edge of the SPB and the hot edge of the $\delta$ Scuti instability strips.
Recently, McNamara et al. (2012) analyzed the light curves of 252 B-star candidates in the Kepler data base
to further characterize
B star's variability and to 
increase the sample of variable B stars for future study.
They classified stars as either constant light emitters,
$\beta$ Cep stars, SPB stars, hybrid pulsators, binaries or stars whose
light curves are dominated by rotation,  hot subdwarfs, or white dwarfs.
Mowlavi et al. (2013) in the case of open cluster NGC 3766 found large population of new variable stars between the red end of SPB type stars and the blue
end of the $\delta$ Scuti type stars where no pulsations are expected to occur on the basis of standard models.
They argued that pulsation could be one of the reasons for showing brightness variation in these stars.

Yang et al. (2013) also presented CCD time series photometric observations for the stars in the open 
clusters NGC 7209, NGC 1582 and Dolidize 18 and found only one star which could be B-type pulsating star.
Jerzykiewicz et al. (2003) presented results of a search for variable stars in the field of the young open cluster NGC 2169 and
found two $\beta$ Cep stars and other type stars.
Diago et al. (2008) have also detected absorption-line B and Be stars in the SMC showing short period variability.
Briquet et al. (2001) studied B-type star HD 131120 and found that this star is monoperiodic with a period of 1.569 d.  They interpreted the variability of this star in terms of a non-radial g-mode pulsation model as well as in terms of a rotational modulation model. They found that rotational modulation model was able to explain the observed line profile variations of the star. 
Luo et al. (2012) carried out $BV$ CCD photometric observations of the open cluster NGC 7654, a young open cluster located in the Cassiopeia constellation, to search for variable stars
and detected 18 SPB stars. They find that the multi-mode
pulsation is more common in the upper part of the MS and g-mode MS pulsating variables probably follow a common period-luminosity relation.

In the light of above discussions we aimed to search for variable stars in the upper part of MS of NGC 1893.
The NGC 1893 cluster is a young open cluster which is associated with nebulae and obscured by dust clouds.
Detailed study of the NGC 1893 on the basis of photometric data has been carried out by Sharma et al. (2007) and Pandey et al. (2013).
 Marco \& Negueruela (2002a) have found the presence of Herbig Ae/Be in the vicinity of O-type stars and B-type MS stars of later spectral type. 
On the basis of spectroscopic survey in the region of NGC 1893 Marco \& Negueruela (2002b) suggest that both Herbig Be stars and classical Be stars are present in NGC 1893. 
Zhang et al. (2008) and Lata et al. (2012) identified a few B-type variable
stars in the cluster NGC 1893.
The previous studies show that NGC 1893 is one of the richest clusters to study the variable stars. 
This paper is in continuation of our efforts to study the MS variable stars in young clusters. In our earlier papers (Lata et al. 2011, 2012), we have presented time series photometry of PMS variable candidates.

The observations of NGC 1893 in the $V$ band have been carried out on 16 nights during December 2007 to January 2013 in order to identify and characterize the variable stars in the NGC 1893 region. In Section 2 we describe the observations and data reduction procedure.  
In Section 3 we discuss the cluster membership of detected variables using $(U-B)/(B-V)$, $(J-H)/(H-K)$ two colour diagram (TCD) and $V/(V-I)$ CMD. 
Section 4 deals with period determination of variables.
Section 5 describes luminosity and temperature of the stars.
The variability characteristics of the stars are discussed in Section 6. 
We discuss X-ray luminosity and period for the variables in Section 7.
Finally, we summarise our results in Section 8.  

\begin{figure*}
\includegraphics[width=17cm]{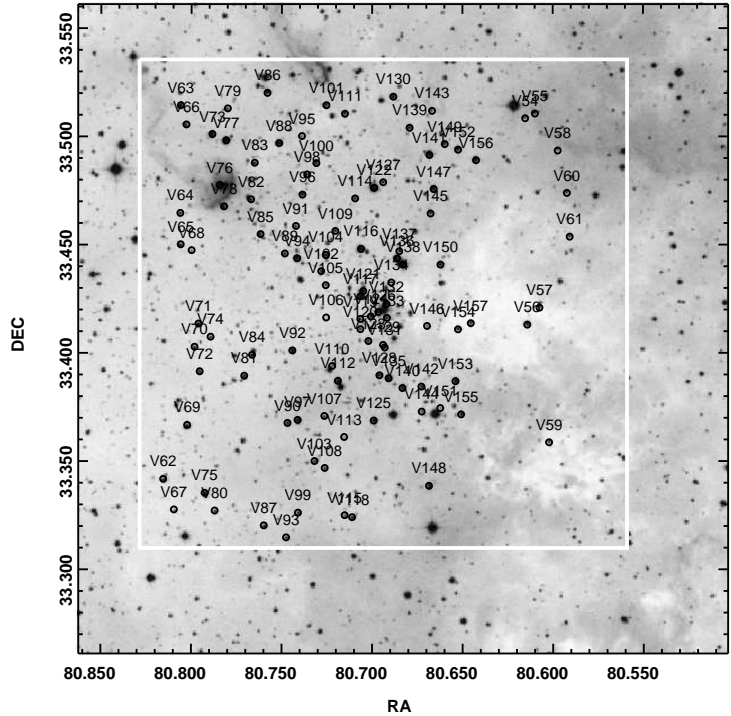}
\caption{ The image of NGC 1893 of 18$\times$18 arcmin$^2$ taken from the DSS-R. The observed region from 104 cm is shown by rectangle. The variable candidates detected in the present work are encircled and labeled with numbers. }
\end{figure*}
%

\setcounter{figure}{1}
\begin{figure*}
\includegraphics[width=18.cm, height=12cm]{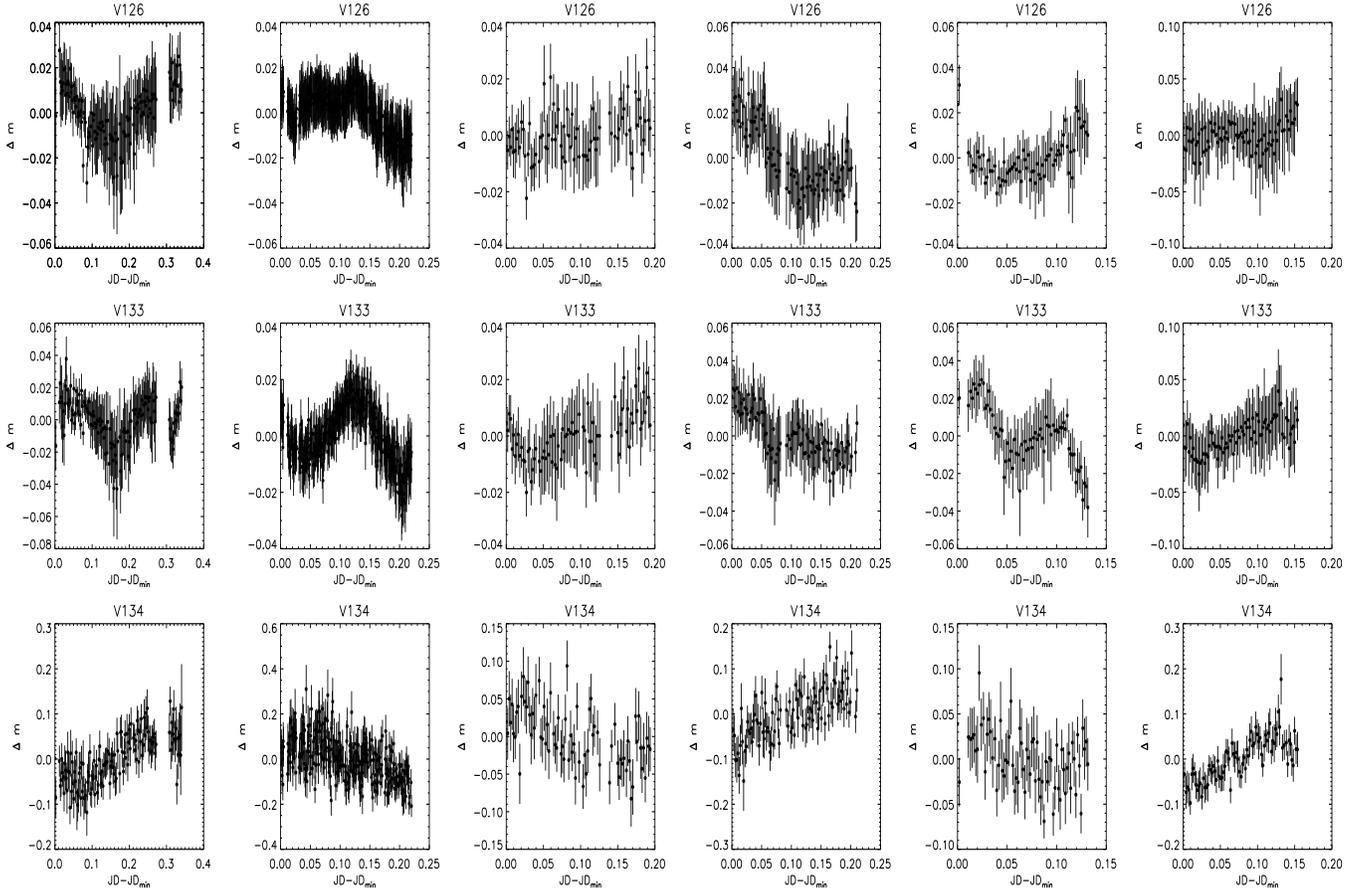}
\caption{Sample light curves of a few variable stars identified in the
present work. The $\Delta$ m represents the differential magnitude in the sense that variable minus comparison star. The complete figure is available online.}
\end{figure*}


\begin{table}
\caption{Log of the observations. N and Exp. represent number of frames obtained and exposure time respectively. \label{tab:obsLog}}
\begin{tabular}{lclcc}
\hline
S. No.&Date of        &Object&{\it V}                &{\it I}                               \\
         &observations&         &(N$\times$Exp.)&(N$\times$Exp.) \\
\hline
1 & 05 Dec 2007 & NGC 1893&3$\times$40s &- \\
2 & 08 Dec 2007 & NGC 1893&3$\times$50s &-\\
3 & 07 Jan 2008 & NGC 1893&2$\times$40s &-\\
4 & 10 Jan 2008 & NGC 1893&3$\times$50s &-\\
5 & 12 Jan 2008 & NGC 1893&80$\times$50s &-\\
6 & 14 Jan 2008 & NGC 1893&70$\times$40s &-\\
7 & 29 Oct 2008 & NGC 1893&97$\times$50s &-\\
8 & 21 Nov 2008 & NGC 1893&137$\times$50s &2$\times$50s\\
9 & 27 Jan 2009 & NGC 1893&5$\times$50s &-\\
10& 28 Jan 2009 & NGC 1893&5$\times$50s &-\\
11& 19 Feb 2009 & NGC 1893&5$\times$50s &-\\
12& 20 Feb 2009 & NGC 1893&3$\times$50s &5$\times$50s\\
13& 20 Feb 2009 & SA 98     &5$\times$90s &5$\times$60s\\
14& 31 Oct 2010 & NGC 1893&3$\times$50s &-\\
15& 22 Dec 2012 & NGC 1893&323$\times$50s &-\\ 
16& 05 Jan 2013 & NGC 1893&80$\times$50s &-\\
\hline
\end{tabular}

\end{table}
\setcounter{figure}{2}
\begin{figure}
\includegraphics[width=9cm]{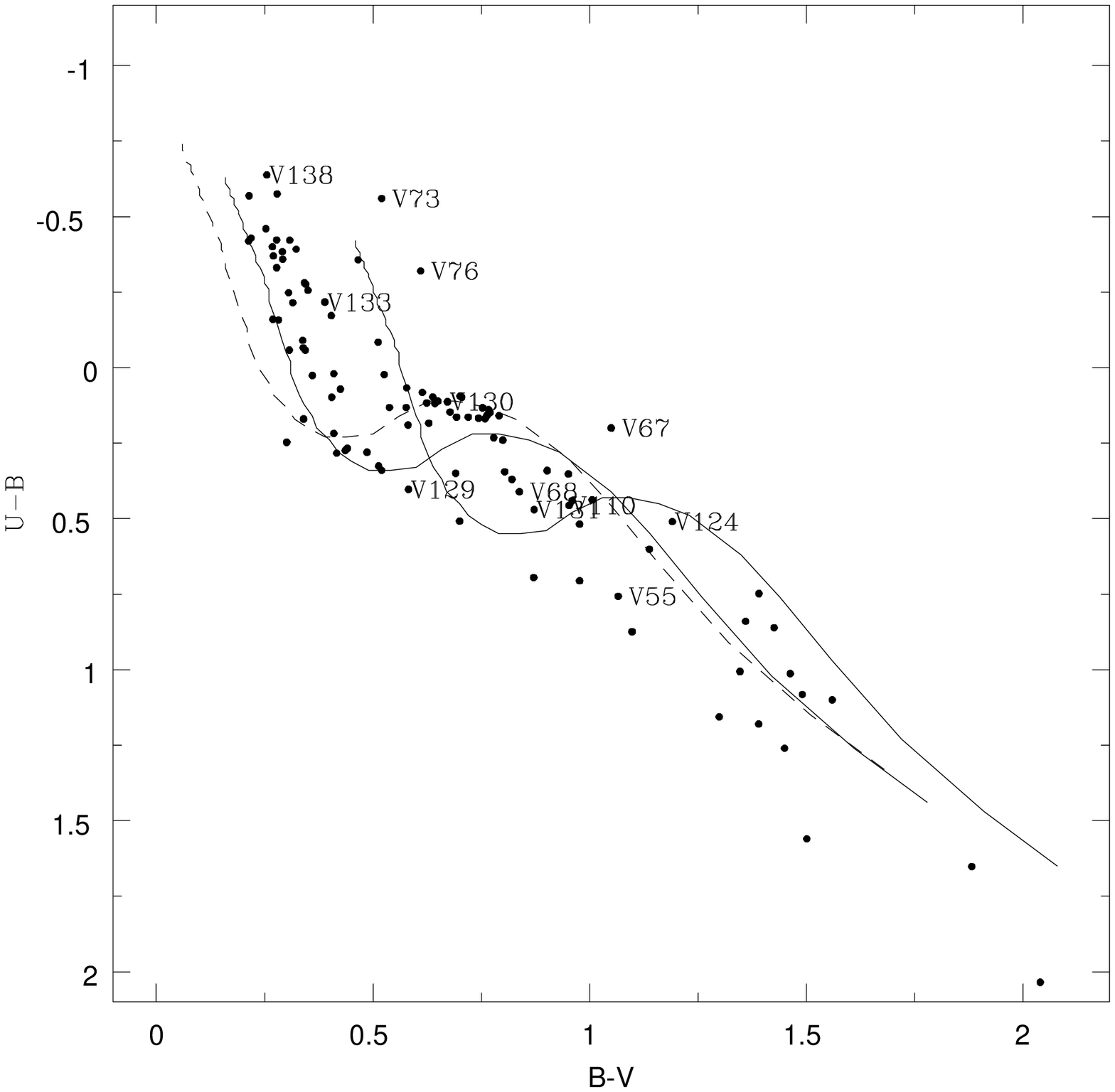}
\caption{$U-B/B-V$ two colour diagram variable stars. The $UBV$ data have been taken from 
Sharma et al. (2007) and Massey  et al. (1995). The solid curve represents the ZAMS by Girardi et al. (2002) shifted along the reddening vector of 0.72 for $E(B-V)_{min}$= 0.4 mag and $E(B-V)_{max}$= 0.7 mag. The dashed curve represents ZAMS by Girardi et al. (2002) for the foreground field population having $E(B-V)$= 0.30 mag. The stars labeled are discussed in section 6.}
\end{figure}

\section{Observations and Data Reduction}
The present work uses observations made from two telescopes; 104-cm ARIES telescope and 130-cm Devasthal telescope.
The details of these two telescopes and instrument are given below.
The photometric observations of the NGC 1893 region were carried out in the
$V$-band on 15 nights and in the $I$-band on two nights during
2007 December 05 to 2013 January 05 using a 2048$\times$2048 CCD camera
attached to the 104-cm Sampurnanand ARIES telescope at Nainital. The field of view is $\sim$13$^{\prime}\times$13$^{\prime}$
and the scale is $\sim0.76^{\prime\prime}$/pixel in 2$\times$2 pixel binning mode. The central position on
the sky was close to RA (2000) = $05^{h}22^{m}42^{s}$ and Dec (2000) = $+33^{\circ}25^{\prime}00^{\prime\prime}$
for all the frames.

In addition, we have also observed the region in $V$-band on 22 December 2012 using
130-cm Devasthal telescope. The 130-cm Devasthal telescope uses
2048$\times$2048 CCD camera having pixel size of 13.5 $\micron$ mounted at the f/4 Cassegrain focus of the telescope. With 0.54 arcsec per pixel plate scale, the entire chip covers a $\sim$ 18$\times$18 arcmin$^2$ field of view on the sky.
Fig. 1, image taken
from Digital Sky Survey (DSS), displays the observed region of NGC 1893.
The observations of NGC 1893 consist of a total of 824 CCD images in the $V$-band.
Bias and twilight flats were also taken
along with the target field. The log of the observations is given in Table \ref{tab:obsLog}.
\subsection{Photometry and variable identification}
 The preprocessing of the CCD images was performed using the IRAF\footnote{IRAF is distributed by the National Optical Astronomy Observatory, which is operated by the Association of Universities for Research in Astronomy (AURA) under cooperative agreement with the National Science Foundation.} and 
the instrumental magnitude of the stars were obtained using the DAOPHOT package (Stetson 1987). 
The details of the procedure can be found in our earlier work (Lata et al. 2012).
We have considered only those stars for further study which have at least 100 observations.
The variable candidates were identified by inspecting their light curves (Sariya, Lata \& Yadav 2014).
The observations taken during December 2007 to October 2010 have already been used in our previous study (Lata et al. 2012) to identify the PMS variable
stars. Present study focuses mainly on the MS stars.
We have identified 104 new variable candidates.
The identification number, coordinates and present $VI$ photometric data for these variable stars
are given in Table 2.  The identification number is in continuation of previous work (Lata et al. 2012).
Sample light curves of a few variables are  shown in Fig. 2. The complete Fig 2. is available in electronic
form.
Table 2 also lists $NIR$ data taken from Prisinzano et al. (2011) and 2MASS catalogue (Cutri et al. 2003).
The procedure of standardization has been given in Lata et al. (2012).
The variable candidates identified in the present sample are marked in Fig. 1.

\setcounter{figure}{3}
\begin{figure}
\includegraphics[width=9cm]{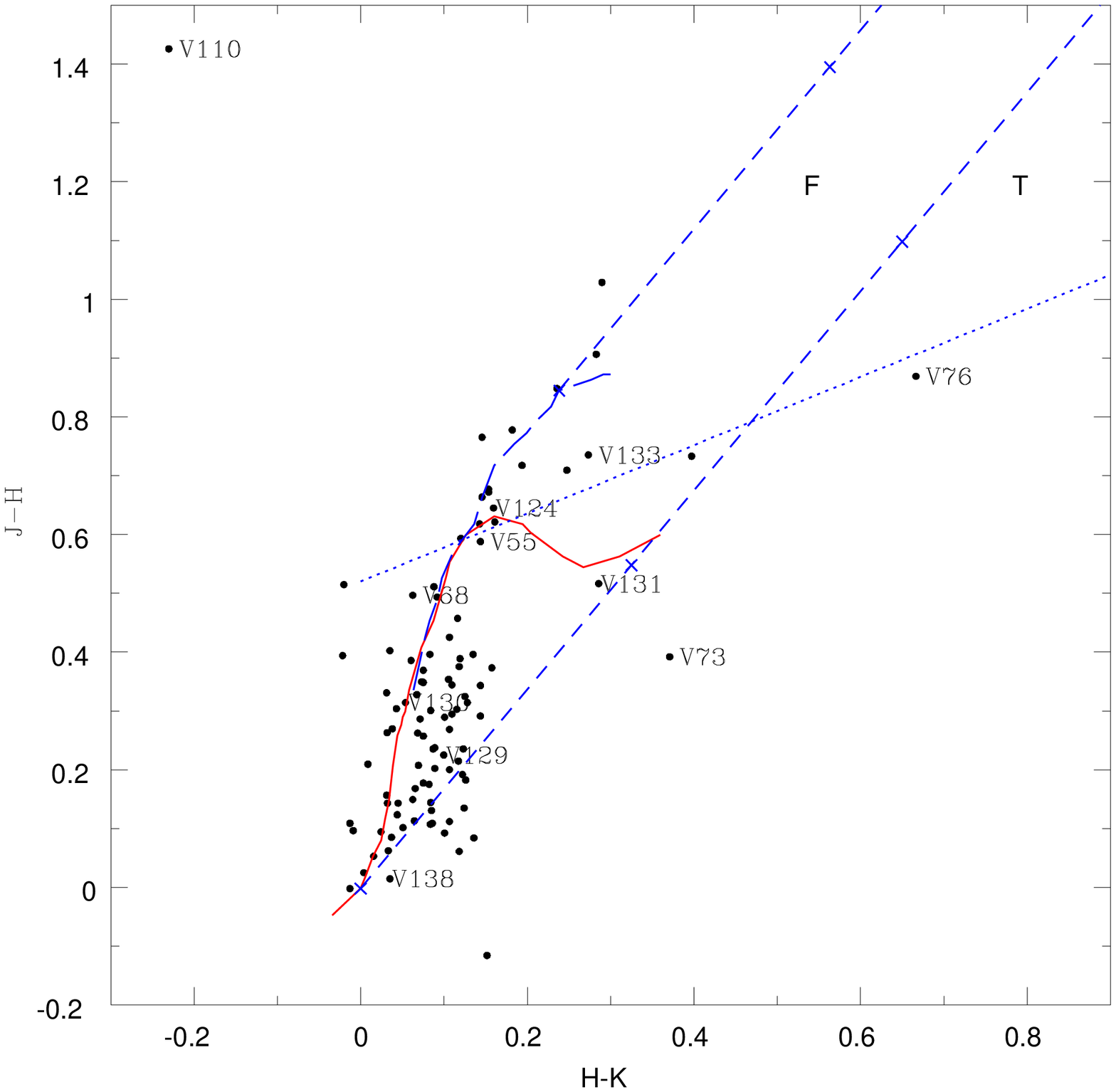}
\caption{($J-H/H-K$) TCD for variable stars lying in the field of
13$\times$13 arcmin of NGC 1893. $JHK$ data have been taken from Prisinzano et al. (2011). The sequences for dwarfs (solid curve) and giants (long dashed curve) are from Bessell \& Brett (1988). The dotted curve represents the locus of TTSs (Meyer et al. 1997).
The small dashed lines represent the reddening vectors (Cohen et al. 1981). 
 The crosses on the reddening vectors represent an increment of visual extinction of $A_{V}$ = 5 mag. The stars labeled are discussed in section 6.} 
\end{figure}

\setcounter{figure}{4}
\begin{figure}
\includegraphics[width=9cm]{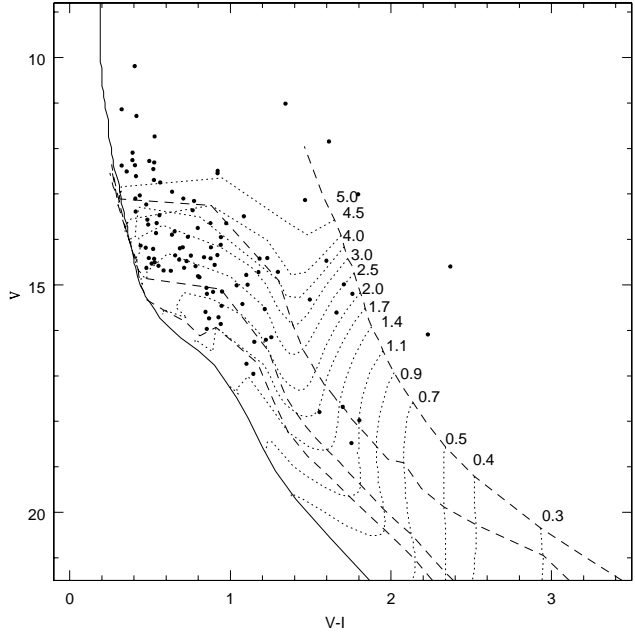}
\caption{$V/V-I$ colour-magnitude diagram of the cluster NGC 1893 for variable candidates . 
The ZAMS by Girardi et al. (2002)
 and PMS isochrones for
0.1, 1, 5, 10 Myrs by Siess et al. (2000) are shown.
The dotted curves show PMS evolutionary tracks of stars of different masses.
The isochrones and evolutionary tracks  are corrected for the cluster distance and $E(V-I)=0.50$ mag. 
}
\end{figure}

\begin{table*}
\caption{The present photometric data, NIR data, period, amplitude and classification of 104 variables in the region of NGC 1893. 
The NIR data have been taken from Prisinzano et al. (2011) and  2MASS point source catalogue (Cutri et al. 2003).
}
\tiny
\begin{tabular}{llllllllllllll}
\hline
ID&
$\alpha_{2000}$&
$\delta_{2000}$&
$V$&
$V-I$&
$J$&
$H$&
$K$&
[Jmko]&
[Hmko]&
[Kmko]&
[period]&
[Amp.]&
Classification. 
\\
&
degree&
degree&
mag&
mag&
mag&
mag&
mag&
mag&
mag&
mag&
day&
mag&

\\
\hline
V54& 80.61530 &33.50847 &  15.489 &   0.975 &     14.006 &     13.622 &    13.524 &  -      &  -      &    -     &   0.211 &  0.015&Field \\
V55& 80.60988 &33.51063 &  14.438 &   1.260 &     12.417 &     11.848 &    11.688 & 13.958 & 13.394 & 13.069  &   0.159 &  0.016&Field\\
V56& 80.61419 &33.41311 &  12.777 &   0.593 &     11.754 &     11.626 &    11.526 &-        &-        & -        &   0.345 &  0.010&Field\\
V57& 80.60744 &33.42097 &  14.617 &   0.813 &     13.279 &     13.050 &    12.948 &-        &-        & -        &   0.236 &  0.009&Field\\
V58& 80.59730 &33.49355 &  15.889 &   0.970 &     14.278 &     13.916 &    13.742 &-        &-        & -        &   0.141 &  0.022&Field\\
V59& 80.60219 &33.35877 &  15.175 &   0.978 &     13.576 &     13.218 &    13.128 &-        &-        & -        &   0.244 &  0.011&Field\\
V60& 80.59224 &33.47397 &  14.531 &   0.559 &     13.617 &     13.485 &    13.345 &-        &-        & -        &   0.236/0.538 &  0.012&MS\\
V61& 80.59069 &33.45375 &  15.183 &   0.922 &     13.708 &     13.370 &    13.280 & 13.698 & 13.395 & 13.261  &   0.211 &  0.012&Field\\
V62& 80.81544 &33.34186 &  13.674 &   1.004 &     12.030 &     11.656 &    11.581 &-        &-        &-         &   0.280 &  0.010&Field\\
V63& 80.80588 &33.51449 &  14.498 &   1.628 &     11.825 &     11.175 &    11.005 &-        &-        &-         &   0.376 &  0.010&PMS\\
V64& 80.80611 &33.46472 &  15.225 &   1.790 &     12.138 &     11.386 &    11.187 & 12.140  &12.027  &13.766   &   0.311 &  0.015&PMS\\
V65& 80.80588 &33.45022 &  14.445 &   0.907 &     13.008 &     12.716 &    12.617 &-        &-        &-         &   0.343 &  0.013&Field\\
V66& 80.80278 &33.50561 &  15.025 &   1.137 &     13.248 &     12.836 &    12.714 &-        &-        &-         &   0.233 &  0.011&Field\\
V67& 80.80950 &33.32772 &  16.281 &   1.179 &         -   &       -     &    -       &-        &-        &-         &   0.079 &  0.020&Field\\
V68& 80.79985 &33.44752 &  16.984 &   1.173 &     15.068 &     14.587 &    14.510 &-        &-        &-         &   0.072 &  0.031&Field\\
V69& 80.80225 &33.36680 &  12.398 &   0.436 &     11.731 &     11.647 &    11.596 &-        &-        &-         &   0.486 &  0.008&MS\\
V70& 80.79819 &33.40291 &  14.506 &   0.757 &     13.183 &     12.902 &    12.786 & 13.193 & 13.216 & 12.745  &   0.535/0.434 &  0.016&MS\\
V71& 80.79602 &33.41363 &  14.438 &   0.522 &     13.627 &     13.423 &    13.401 &-        &-        &-         & 0.520/0.638 &  0.014&MS\\
V72& 80.79533 &33.39155 &  14.212 &   0.503 &     13.258  &    13.105  &   13.060  & 13.518 & 13.281 & 13.054  &   0.209 &  0.011&MS\\
V73& 80.78839 &33.50119 &  12.520 &   0.951 &     10.785 &     10.405 &    10.013 & 10.905 & 10.434 & 10.116  &   0.269 &  0.007&Herbig Ae/Be\\
V74& 80.78932 &33.40755 &  14.224 &   0.715 &     12.985 &     12.699 &    12.574 & 13.109 & 12.887 & 12.666  &   0.249 &  0.008&MS\\
V75& 80.79255 &33.33544 &  13.163 &   1.495 &     10.595 &      9.953 &     9.791 &-        &-        &-         &   0.238 &  0.009&PMS\\
V76& 80.78433 &33.47766 &  14.743 &   1.326 &     11.789 &     10.949 &    10.255 & 14.622 & 13.723 & 13.018  &   0.255 &  0.034&PMS\\
V77& 80.78074 &33.49830 &  11.876 &   1.644 &      9.196 &      8.457 &     8.295 &-        &-        &-         &   0.229 &  0.007&PMS\\
V78& 80.78194 &33.46780 &  13.261 &   0.506 &-            &-            &-           & 12.756 & 12.502 & 12.468  &   0.423/0.655/0.575 &  0.019&MS\\
V79& 80.77986 &33.51302 &  14.712 &   0.614 &-            &-            &-           & 14.220  &13.756  &13.783   &   0.290 &  0.017&MS\\
V80& 80.78700 &33.32725 &  15.637 &   1.691 &     12.689 &     11.995 &    11.784 & 14.172 & 14.237 & -       &   0.233 &  0.015&PMS\\
V81& 80.77069 &33.38955 &  14.165 &   0.471 &-            &-            &-           &-        &-        &-         &   0.177 &  0.007&MS\\
V82& 80.76699 &33.47105 &  13.670 &   0.571 &     12.771 &     12.574 &    12.470 & 16.247 & 15.334 & 14.684  &   0.510 &  0.006&MS\\
V83& 80.76494 &33.48794 &  13.890 &   0.567 &     13.052 &     12.991 &    12.857 &-        &-        &-         &   0.458/0.478 &  0.007&MS\\
V84& 80.76638 &33.39919 &  14.655 &   0.507 &     -       &-            &-           &-        &-        &-         &   0.310/0.354 &  0.010&MS\\
V85& 80.76171 &33.45497 &  14.203 &   0.909 &     -       &-            &-           &-        &-        &-         &   0.287 &  0.006&Field\\
V86& 80.75791 &33.52019 &  13.037 &   1.826 &     10.296 &      9.641 &     9.471 &  -     &  -     & -       &   0.246 &  0.010&PMS\\
V87& 80.75986 &33.32041 &  14.625 &   2.400 &     10.359 &      9.365 &     9.056 &  -      &  -      &-         &   0.260 &  0.011&PMS\\
V88& 80.75141 &33.49702 &  12.337 &   0.556 &     11.636 &     11.529 &    11.428 &  -      &  -      &-         &   0.466/0.389 &  0.013&MS\\
V89$^{*}$& 80.74827 &33.44599 &  14.238 &   0.549 &     13.282 &     13.199 &    13.047 &  -      &  -      &-         &   0.586 &  0.016&MS\\
V90& 80.74678 &33.36774 &  14.653 &   0.742 &      13.700 &     13.529 &    13.432 & -       & -       &-         &   0.279/0.254 &  0.009&MS\\
V91$^{*}$& 80.74214 &33.45872 &  14.559 &   0.540 &      12.395&     12.255 &     12.209&  -        &-        &-         &   0.511/0.576 &  0.016&MS\\
V92$^{*}$& 80.74407 &33.40124 &  13.130 &   0.438 &     12.395 &     12.255 &    12.209 &  13.468&  13.178&  13.038 &   0.594/0.474 & 0.016&MS\\
V93& 80.74752 &33.31486 &  14.838 &   0.829 &     13.421 &     13.192 &    13.053 &-        &-        &-         &   0.327 &  0.012&Field\\
V94$^{*}$& 80.74150 &33.44383 &  12.122 &   0.421 &      11.460 &      11.32 &    11.261 &  11.781&  12.814&  11.746 &   0.603/0.376/0.670/0.875 &  0.007&MS\\
V95& 80.73883 &33.50027 &  15.623 &   0.875 &    14.175  &    13.966  &   13.833  &-        &-        &-         &   0.194 &  0.012&Field\\
V96& 80.73855 &33.47322 &  13.520 &   1.114 &     11.758 &     11.263 &    11.160 &  11.640 & 11.333 & 11.101  &   0.250 &  0.006&Field\\
V97& 80.74116 &33.36914 &  12.535 &   0.384 &     11.911 &     11.816 &    11.812 &-        &-        &-         &   0.483/3.75 &  0.007&MS\\
V98& 80.73594 &33.48255 &  14.719 &   0.659 &     13.668 &     13.558 &    13.436 &-        &-        &-         &   0.458/0.478/0.622 &  0.019&MS\\
V99& 80.74097 &33.32630 &  15.225 &   0.884 &     13.738 &     13.444 &    13.313 &-        &-        &-         &   0.179 &  0.013&Field\\
V100& 80.73085 &33.48777 &  14.451 &   1.212 &     12.478 &     12.094 &    11.943 &-        &-        &-         &   0.137 &  0.007&Field\\
V101& 80.72528 &33.51449 &  15.017 &   1.737 &     12.181 &     11.472 &    11.053 &  12.097&  11.482&  11.143 &   0.245 &  0.010&PMS\\
V102& 80.72816 &33.43786 &  14.380 &   0.686 &     13.235 &     13.040 &    12.918 &-        &-        &-         &   0.424/0.633 &  0.011&MS\\
V103& 80.73185 &33.35013 &  15.560 &   1.244 &     13.418 &     12.844 &    12.708 &-        &-        &-         &   0.303 &  0.014&Field\\
V104$^{*}$& 80.72563 &33.44538 &  13.417 &   0.440 &     12.798 &     12.677 &    12.619 &  -     &  12.615&  12.509 &   0.476/0.653 &0.006&MS\\
V105& 80.72566 &33.43138 &  15.765 &   0.897 &     14.261 &     13.871 &    13.822 &-        &-        &-         &   0.294 &  0.013&Field\\
V106& 80.72541 &33.41644 &  18.507 &   1.784 &     15.149 &     14.463 &    14.197 &  15.136&  14.475&  14.099 &   0.187 &  0.079&PMS\\
V107& 80.72641 &33.37097 &  14.861 &   0.837 &     13.434 &     13.113 &    13.068 &-        &-        &-         &   0.268 &  0.009&Field\\
V108& 80.72621 &33.34694 &  15.449 &   1.105 &     13.524 &     13.191 &    13.031 &  15.781&  15.704&  15.244 &   0.179 &  0.014&Field\\
V109$^{*}$& 80.72033 &33.45644 &  14.609 &   0.582 &     13.681 &     13.376 &    13.232 &  13.685&  13.424&  13.277 &   0.511/0.623 &  0.014&MS\\
V110& 80.72211 &33.39405 &  16.764 &   1.130 &-            &-            &-           &  15.961&  14.737&  15.011 &   0.068 &  0.039&Field\\
V111& 80.71502 &33.51058 &  13.979 &   0.971 &     12.425 &     12.164 &    12.042 &-        &-        &-         &   0.224 &  0.007&Field\\
V112$^{*}$& 80.71888 &33.38719 &  12.405 &   0.353 &     11.822 &      11.76 &    11.713 &-        &-        &-         &   0.325 &  0.009&MS \\
V113& 80.71544 &33.36125 &  18.006 &   1.833 &    14.776  &    13.956  &   13.702  &  14.645&  13.907&  13.724 &   0.279 &  0.068&PMS\\
V114& 80.70938 &33.47144 &  14.420 &   0.866 &     13.075 &     12.797 &    12.711 &-        &-        &-         &   0.380 &  0.007&Field\\
V115& 80.71516 &33.32513 &  16.180 &   1.285 &      13.950 &     13.507 &    13.375 &-        &-        &-         &   0.354 &  0.020&Field\\
V116$^{*}$& 80.70619 &33.44816 &  11.168 &   0.353 &     10.716 &      10.69 &    10.673 &-        &-        &-         &   0.505/0.616/0.455 &  0.015&MS\\
V117& 80.70655 &33.42625 &  12.722 &   0.554 &     11.806 &     11.665 &    11.566 &  12.128&  12.254&  11.851 &   0.596 &  0.018&MS\\
V118& 80.71108 &33.32419 &  14.148 &   0.970 &     12.479 &     12.145 &    12.020 &-        &-        &-         &   0.182 &  0.012&Field\\
V119& 80.70652 &33.41580 &  15.039 &   1.069 &     13.238 &     12.760 &    12.653 &-        &-        &-         &   0.269 &  0.011&Field\\
V120& 80.70647 &33.41105 &  15.093 &   0.881 &     13.627 &     13.263 &    13.129 &  14.277&  13.884&  13.722 &   0.209 &  0.012&Field\\
V121& 80.70488 &33.42866 &  13.702 &   0.518 &     -  &      -  &     -  &  13.197&  13.008&  12.555 &   0.448 &  0.009&MS\\
V122& 80.69889 &33.47633 &  11.043 &   1.373 &      8.816 &      8.218 &     8.059 &-        &-        &-         &   0.195 &  0.008&Field\\
V123& 80.70205 &33.40558 &  14.454 &   0.555 &     13.428 &     13.337 &    13.221 &  13.782&  14.053&  13.376 &   0.190 &  0.008&MS\\
V124$^{**}$& 80.70066 &33.41694 &  15.351 &   1.525 &     12.771 &     12.147 &    11.971 &  13.865&  12.701&  12.327 &   0.242 &  0.015&PMS\\
V125& 80.69908 &33.36886 &  12.574 &   0.950 &     10.921 &     10.582 &    10.494 &  -     &  -     &  -      &   0.436 &  0.008&MS\\
V126& 80.69660 &33.41894 &  13.061 &   0.466 &     -       &-            &    -       &  12.439&  12.501&  12.326 &   0.501/0.416/0.732 &  0.010&MS\\
V127& 80.69386 &33.47894 &  14.373 &   0.949 &     12.902 &     12.584 &    12.502 & -       &-        & -        &   0.177 &  0.007&Field\\
V128$^{*}$&80.69594 &33.38975 &  13.851 &   0.683 &     12.578 &     12.263 &    12.122 &  12.632&  12.275&   12.160 &   0.269 &  0.007&MS\\
V129$^{**}$&80.69394 &33.40380 &  14.385 &   0.784 &     12.938 &     12.719 &    12.604 &  13.150 & 13.145 & 12.655  &   0.309 &  0.009&MS\\
V130& 80.68824 &33.51844 &  13.671 &   0.906 &     12.269 &     11.964 &    11.896 &  12.173&  11.885&  11.802 &   0.211 &  0.010&Field\\
V131$^{**}$&80.69300 &33.40258 &  14.748 &   1.205 &     12.604 &     12.104 &    11.799 &  12.761&  12.442&  11.857 &   0.208 &  0.010&PMS\\
V132& 80.69205 &33.42266 &  12.482 &   0.550 &       -     &  -          &  -         &  11.485&  11.361&  11.266 &   0.554/0.570 &  0.008&MS\\
V133$^{*}$& 80.69178 &33.41624 &  13.925 &   0.666 &       -     &  -          &  -         &  14.571&  13.923&  13.605 &   0.485/0.467 &  0.012&MS\\
V134& 80.68939 &33.43264 &  17.825 &   1.585 &     11.885 &     11.778 &    11.778 &  14.819&  14.194&  13.989 &   0.376 &  0.067&Field\\
V135$^{*}$& 80.69088 &33.38838 &  12.638 &   0.442 &     10.703 &      10.65 &    10.621 &  -      &  -      & -        &   0.353 &  0.007&MS\\
V136$^{*}$& 80.68619 &33.44366 &  11.315 &   0.445 &    -        &     -       & -          &  10.545&  10.508&  10.628 &   0.541/0.429/0.646 &  0.008&MS\\
V137& 80.68497 &33.44708 &  16.236 &   1.251 &    -        &     -       & -          &  -      & -       &   -      &   0.382 &  0.021&Field\\
V138$^{*}$& 80.68297 &33.44086 &  10.218 &   0.434 &     9.648  &     9.633  &    9.584  &  -     &   9.394&   9.532 &   0.570/0.593/0.456 &  0.006&MS\\
V139& 80.67925 &33.50408 &  13.778 &   0.827 &     12.481 &     12.226 &    12.143 &   -     &    -    &  -       &   0.411/0.333 &  0.015&MS\\
V140$^{*}$& 80.68321 &33.38386 &  13.595 &   0.513 &     12.619 &     12.446 &    12.356 &  12.612&  12.383&  12.338 &   0.414 &  0.008&MS\\
V141& 80.66828 &33.49152 &  11.765 &   0.558 &     10.932 &     10.821 &    10.742 &   -     &    -    &   -      &   0.477 &  0.006 &Field\\
V142$^{*}$& 80.67275 &33.38455 &  14.473 &   0.711 &     13.179 &     13.001 &    12.859 &  13.154&   12.92&  12.902 &   0.210 &  0.009&MS\\
V143& 80.66683 &33.51191 &  15.740 &   0.956 &     14.227 &     13.845 &    13.854 &  14.143&  13.813&  13.746 &   0.143 &  0.017&Field\\
V144& 80.67258 &33.37300 &  16.119 &   2.260 &     12.068 &     11.192 &    10.890 &  -      &  -      &  -       &   0.320 &  0.017&PMS\\
V145& 80.66766 &33.46450 &  13.973 &   0.765 &     12.766 &     12.535 &    12.431 &  -      &  -      &  -       &   0.410/0.470/0.564 &  0.010&MS\\
V146& 80.66963 &33.41255 &  15.996 &   0.883 &     14.451 &     14.074 &    13.939 &  -      &  -      &  -       &   0.297 &  0.017&Field\\
V147& 80.66594 &33.47577 &  13.183 &   0.804 &     11.918 &     11.623 &    11.566 &  -      &  -      &  -       &   0.369 &  0.006&Field\\
V148& 80.66855 &33.33869 &  13.388 &   0.795 &     12.007 &     11.757 &    11.667 &  -      &  -      &  -       &   0.283 &  0.010&Field\\
V149& 80.65980 &33.49655 &  14.589 &   0.932 &       13.100 &     12.757 &  12.636 &  -      &  -      &  -       &   0.179 &  0.010&Field\\
V150$^{*}$& 80.66217 &33.44088 &  13.130 &   0.737 &     11.894 &     11.707 &    11.569 &  11.902&  11.711&  11.628 &   0.415/0.440/0.520 &  0.008&MS\\
V151& 80.66233 &33.37463 &  17.712 &   1.730 &     -       &    -        &    -       &  15.119&  14.566&   14.380 &   0.528 &  0.072&PMS\\
V152& 80.65242 &33.49397 &  12.285 &   0.420 &     11.635 &     11.535 &     11.470 &  -      &   -     &  -       &   0.502 &  0.006&Field\\
V153$^{*}$& 80.65391 &33.38702 &  12.981 &   0.667 &     11.839 &     11.693 &    11.616 &  -      &   -     &  -       &   0.544 &  0.019&MS\\
V154& 80.65252 &33.41099 &  13.496 &   0.588 &     12.427 &     12.263 &    12.183 &  -      &   -     &  -       &   0.258 &  0.008&Field\\
V155& 80.65080 &33.37169 &  14.810 &   1.129 &     12.793 &     12.510 &     12.350 &  -      &   -     &  -       &   0.247 &  0.010&Field\\
V156$^{*}$& 80.64252 &33.48908 &  12.305 &   0.525 &     11.526 &     11.433 &    11.395 &  -      &   -     &  -       &   0.479 &  0.006&MS\\
V157& 80.64544 &33.41380 &  14.203 &   0.733 &     12.985 &     12.723 &    12.671 &  -      &   -     &  -       &   0.294 &  0.011&MS\\
\hline
\end{tabular}
${^*}$:O/B/Be type stars from Marco \& Negueruela (2002a, b) \\
$^{**}$: PMS stars from Marco \& Negueruela (2002a, b) \\
\end{table*}

\section{Membership}
The $(U-B)/(B-V)$ and $(J-H)/(H-K)$ TCD, and $V/(V-I)$ CMD have been used to find out
the association of the identified variables with the cluster NGC 1893 and to know the nature of the present variable candidates.

\subsection{$(U-B)/(B-V)$ and $(J-H)/(H-K)$ two colour Diagrams}
The $U-B/B-V$ TCD for variable candidates is shown in Fig. 3.  
The $UBV$ data for 92 common variable candidates have been taken from 
Sharma et al. (2007).
Since present sample consists of 104 stars, for remaining 12 stars $UBV$ data have been taken from Massey et al. (1995). The distribution of stars in the $(U-B)/(B-V)$ TCD indicates a variable reddening in the cluster region with $E(B-V)$ from $\sim$0.4 to 0.7 mag. 
The majority of the stars having $E(B-V)$ from $\sim$0.4 to 0.7 mag have 
 $(B-V)$ colours $\lesssim$ 0.6 mag.  
These could be OB-population of the NGC 1893 cluster. 
To further classify these stars we used $(J-H)/(H-K)$ TCD as shown in Fig. 4.
The $JHK$ data have been taken from Prisinzano et al. (2011) and 2MASS.
Out of 104 variables 98 stars have  their $JHK$ counter parts. 
The $JHK$ data catalogued by Prisinzano et al. (2011) are in MKO system which 
 were converted to 2MASS system using relations given on the website\footnote[2]{http://www.astro.caltech.edu/~jmc/2mass/v3/transformations/}. After that 
$JHK$ data from both the catalogues (Prisinzano catalogue and 2MASS catalogue) were transformed to CIT system using the relations given on the above mentioned website. The solid
and long dashed lines in Fig. 4 represents unreddened MS and giant loci (Bessell \& Brett 1988) respectively.
The dotted line indicates the intrinsic locus of classical T-tauri stars (CTTSs) (Meyer et al. 1997).
The parallel dashed lines are the reddening vectors drawn from the tip (spectral type M4) of the giant branch (`left reddening line'), from the base (spectral type A0) of the main-sequence branch (`middle reddening line') and from the tip of the intrinsic CTTS line (`right reddening line'). The extinction ratios $A_J/A_V= 0.265, A_H/A_V= 0.155$ and $A_K/A_V= 0.090$ have been adopted from Cohen et al. (1981).
The sources lying in `F' region could be
either field stars (MS stars, giants) or Class III and Class II sources
with small NIR excesses. The sources lying in the `T' region are considered to be
mostly CTTSs (Class II objects). The
$(J-H)/(H-K)$ TCD shows that the majority of the variable candidates are located below TTS locus and these should be MS stars. 

\subsection {$V/V-I$ colour-magnitude diagram}
The $V/V-I$ CMD for all the variable candidates have been plotted  in Fig 5. In
Fig. 5 we have also plotted theoretical isochrone for 4 Myr for Z=0.02 (continuous line) by Girardi et al. (2002) and PMS isochrones for various ages as well as evolutionary tracks for various masses by
Siess et al. (2000). All the isochrones and evolutionary tracks are corrected for the distance (3.25 kpc) and minimum reddening $E(V-I)$=0.50 mag. The minimum value of $E(V-I)$ has been estimated using the relation $E(V-I)/E(B-V)=1.25$ and $E(B-V)=0.40$ mag.

The identified variables are classified as MS, PMS and field stars on the basis of their location in the $(U-B)/(B-V)$, $(J-H)/(H-K)$ TCDs and $V/(V-I)$ 
CMD and the classification is given in the 14th coloumn of Table 2. 

\setcounter{figure}{5}
\begin{figure}
\includegraphics[width=9cm, height=8cm]{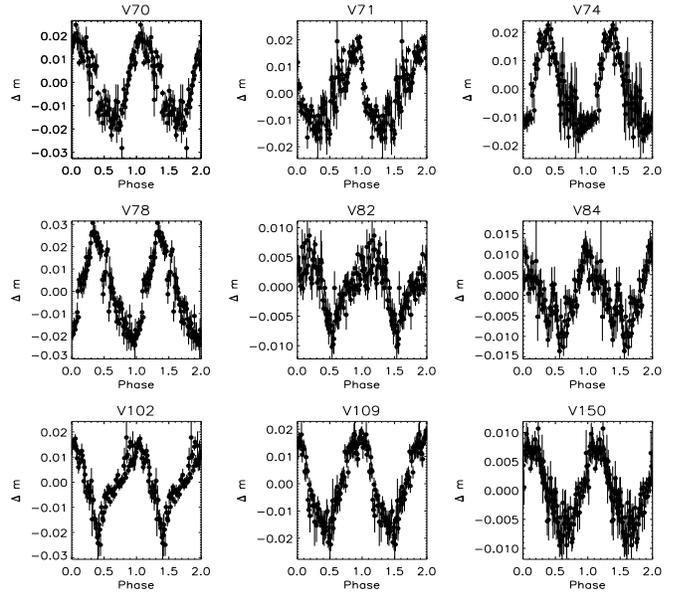}
\caption{The sample phased light curves of MS variable stars. The complete figure is available electronic
form only.}
\end{figure}

\section{Period Determination}
We used the Lomb-Scargle (LS) periodogram (Lomb 1976; Scargle 1982) to determine the most probable
period of a variable star. The LS method is useful to estimate 
periodicities even in the case of unevenly spaced data. We used the algorithm
available at the Starlink\footnote[3]{http://www.starlink.uk} software 
database.
The periods were further verified with the software period04\footnote[4]{http://www.univie.ac.at/tops/Period04} (Lenz \& Breger 2005). 
The software period04 provides the frequency and semi-amplitude of the variability in a light curve.
Periods derived from the LS method and Period04 generally
matched well.
The most probable periods with amplitude are listed in Table 2.
The light curves of variable stars are folded with their estimated period.
Sample phased light curves of identified MS, PMS and field stars are shown
in Figs. 6, 7 and 8 respectively, where
averaged differential magnitude 
in 0.01 phase bin along with $\sigma$ error bars have been plotted. The phased light curves of all the MS, PMS and field variable stars are
available online.

\setcounter{figure}{6}
\begin{figure}
\includegraphics[width=9cm, height=8cm]{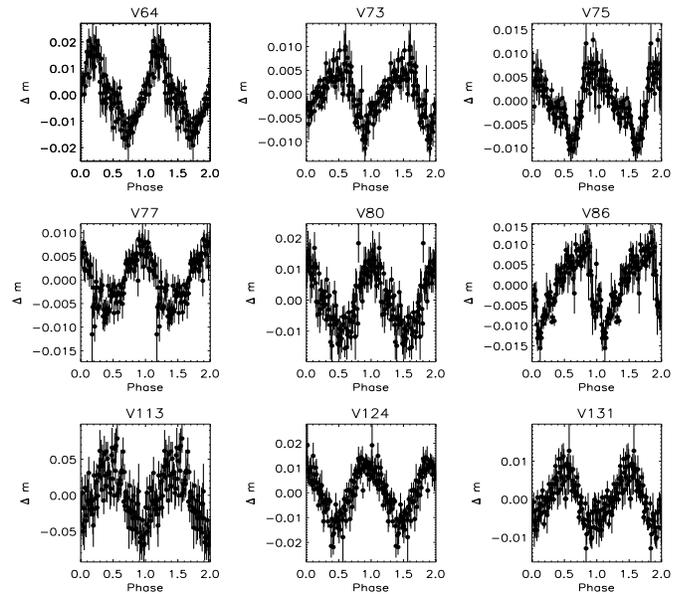}
\caption{Same as Fig. 6 but for PMS variable stars. The complete figure is available electronic
form only.}
\end{figure}

\section{$M_{V}/(B-V)_{0}$ and Luminosity $(L/L_{\odot}$) vs. Effective Temperature ($T _{eff}$) diagram}
The $M_{V}/(B-V)_{0}$ CMD for the identified MS and PMS stars is shown in Fig. 9. The intrinsic $(B-V)$ colours of MS stars have been
determined using Q-method as described by Gutierrez-Moreno (1975).
For PMS stars we have used average reddening ($E(B-V)$=0.55 mag) of the region. Fig. 10 shows $\log(L/L_{\odot})/\log T_{eff}$ diagram for the MS variables.
The absolute magnitude $M_{V}$ was converted to luminosity using the relations $\log(L/L_{\odot})=-0.4(M_{bol}-M_{bol,\odot})$, and $M_{bol}=M_{V}+BC$, where BC is
the bolometric correction. The bolometric magnitude $M_{bol,\odot}$ for the Sun has been taken 4.73 mag (Torres 2010). 
To determine BC and effective temperature $T_{eff}$ 
we used the relations between $T_{eff}$-  
intrinsic $(B-V)$ colours, and between  $T_{eff}$-BC by Torres (2010). 
The luminosity ($\log L/L_{\odot}$), $M_{bol}$ and $\log T_{eff}$ and BC of MS stars
are listed in Table 3.

\begin{table}
\caption{The effective temperature $T_{eff}$, bolometric correction (BC),  bolometric magnitude ($M_{bol}$), luminosity (L) and classification for MS stars.}
\scriptsize
\begin{tabular}{llccll}
\hline
ID&$\log T_{eff}$&   BC &  $M_{bol}$ & $\log(L/L_{\odot})$&Classification \\
\hline
         V60   &    3.988 &   -0.03125  &    0.7067  &     1.610 & new  \\
         V69   &    4.203 &   -1.35900  &   -2.6790  &     2.965 & SPB  \\
         V70   &    3.987 &   -0.07031  &    0.6997  &     1.613 & new  \\
         V71   &    4.092 &   -0.60940  &    0.0886  &     1.857 & new  \\
         V72   &    3.997 &   -0.09375  &    0.3563  &     1.750 & new  \\
         V74   &    4.030 &   -0.28120  &    0.1937  &     1.815 & new  \\
         V78   &    4.182 &   -1.16400  &   -1.6810  &     2.565 & SPB  \\
         V79   &    4.058 &   -0.42190  &    0.5581  &     1.670 & new  \\
         V81   &    4.089 &   -0.55470  &   -0.1487  &     1.952 & new  \\
         V82   &    4.071 &   -0.50000  &   -0.5850  &     2.127 & SPB  \\
         V83   &    4.131 &   -0.92190  &   -0.7899  &     2.209 & SPB  \\
         V84   &    4.059 &   -0.54690  &    0.3631  &     1.748 & new  \\
         V88   &    4.260 &   -1.50000  &   -2.9320  &     3.066 & SPB  \\
         V89   &    4.043 &   -0.38280  &    0.0952  &     1.855 & new  \\
         V90   &    4.050 &   -0.46090  &    0.4041  &     1.731 & new  \\
         V91   &    4.080 &   -0.55470  &    0.2533  &     1.791 & new  \\
         V92   &    4.206 &   -1.32800  &   -1.9910  &     2.689 & SPB  \\
         V94   &    4.259 &   -1.63300  &   -3.2830  &     3.206 & SPB  \\
         V97   &    4.208 &   -1.32800  &   -2.5790  &     2.924 & SPB  \\
         V98   &    4.043 &   -0.31250  &    0.6455  &     1.635 & new  \\
        V102   &    3.989 &   -0.02344  &    0.6006  &     1.653 & new  \\
        V104   &    4.223 &   -1.32000  &   -1.6550  &     2.555 & SPB  \\
        V109   &    4.069 &   -0.50000  &    0.3570  &     1.750 & new  \\
        V112   &    4.218 &   -1.42200  &   -2.7830  &     3.006 & SPB  \\
        V116   &    4.328 &   -2.07800  &   -4.6720  &     3.762 & $\beta$ Cep \\
        V117   &    4.179 &   -1.03100  &   -2.0690  &     2.720 & SPB  \\
        V121   &    4.144 &   -0.96880  &   -1.0250  &     2.303 & SPB  \\
        V123   &    3.983 &   -0.06250  &    0.6215  &     1.644 & new  \\
        V125   &    4.078 &   -0.57810  &   -1.7990  &     2.612 & SPB  \\
        V126   &    4.245 &   -1.36700  &   -2.0970  &     2.732 & SPB  \\
        V128   &    4.142 &   -0.84380  &   -0.7797  &     2.205 & SPB  \\
        V129   &    3.983 &    0.06250  &    0.6515  &     1.632 & new  \\
        V132   &    4.168 &   -0.94530  &   -2.2320  &     2.786 & SPB  \\
        V133   &    4.161 &   -1.01600  &   -0.8546  &     2.235 & SPB  \\
        V135   &    4.223 &   -1.59400  &   -2.7500  &     2.993 & SPB  \\
        V136   &    4.379 &   -2.14100  &   -4.6030  &     3.734 & $\beta$ Cep \\
        V138   &    4.428 &   -2.39100  &   -5.9480  &     4.272 & $\beta$ Cep \\
        V139   &    4.041 &   -0.29690  &   -0.2839  &     2.006 & SPB  \\
        V140   &    4.177 &   -1.18000  &   -1.3790  &     2.444 & SPB  \\
        V142   &    4.136 &   -0.81250  &   -0.1195  &     1.941 & SPB  \\
        V145   &    4.061 &   -0.43750  &   -0.2515  &     1.993 & new  \\
        V150   &    4.090 &   -0.64840  &   -1.3020  &     2.414 & SPB  \\
        V153   &    4.301 &   -1.70300  &   -2.5100  &     2.897 & SPB  \\
        V156   &    4.246 &   -1.30500  &   -2.8010  &     3.013 & SPB  \\
        V157   &    4.087 &   -0.71880  &   -0.3117  &     2.017 & SPB  \\
\hline
\end {tabular}
\end{table}

\setcounter{figure}{7}
\begin{figure}
\includegraphics[width=9cm, height=8cm]{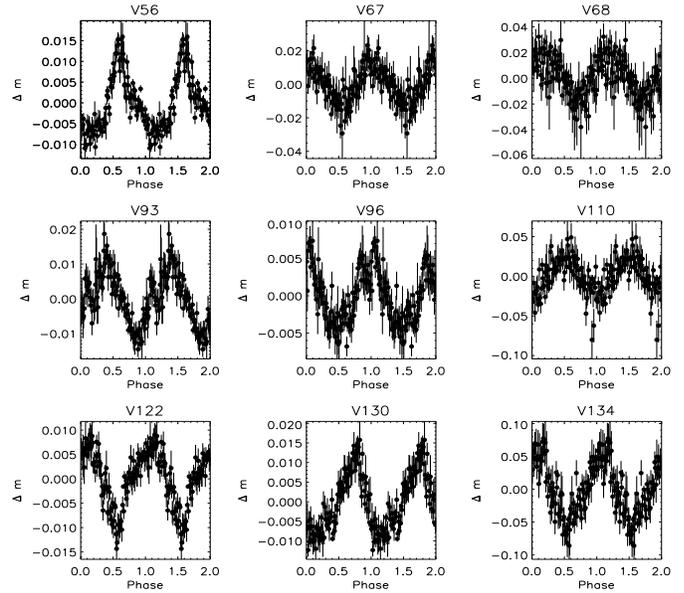}
\caption{Same as Fig. 6 but for field variable stars. The complete figure is available electronic
form only.}
\end{figure}

\section{Variability characteristics}

\subsection{MS variables}
In this section we will characterize the MS variables identified in the present study.
To characterize the variable stars we need period, amplitude and shape of the light curves. Additionally, 
the position of the variable stars in the $H-R$ diagram is also needed 
to ascertain their nature.
Fig. 10 shows the $H-R$ diagram for variables along with the theoretical SPB instability (continuous curve), location of $\beta$ Cep stars (dashed curve) and empirical 
$\delta$ instability strip (dotted curve) taken from Balona et al. (2011; references therein).

Forty five stars are found to be of MS type stars. We find that the observed range
of periods for these stars is between 0.17 to 0.6 d. The amplitude of these stars is of level of a few mmag). 
These could be $\beta$ Cep, SPB and classical Be type stars. 
In the $H-R$ diagram the lower limit of $\log L/L_{\odot}$ for $\beta$ Cep type is predicted as $\gtrsim$ 3, whereas for SPB variables
the lower limit of $\log L/L_{\odot}$ as $\sim$ 1.8. The $\delta$ Scuti type stars have upper limit as $\log L/L_{\odot}$=1.
There is about 2 mag gap between red end of SPB type stars and blue end of $\delta$ Scuti type stars (Mowlavi et al. 2013).
Mowlavi et al. (2013) analyzed the population of periodic variable stars in the open cluster NGC 3766.
They found
a large population of new variable stars between SPB stars and the $\delta$ Scuti stars, the region
where no pulsations were predicted based on the standard stellar models. The periods of these variable stars ranges
from 0.1 to 0.7 d, with amplitudes between 1.0 to 4.0 mmag. They found that 20\% stars are variable in that region
within their detection limit and expected more stars to be variable in this region.
The origin of variability of these stars could be pulsation. One of the probable causes of pulsation in these stars
could be rapid rotation which alter the internal conditions of a star enough to sustain stellar pulsations. The second
cause for the brightness variation in these stars might be the presence of spots on the surface of such rotating stars and that these spots would induce light variations as the star rotates. But hot stars are not expected to be active, and no theory can currently explain how spots could be produced on the surface of such stars.
Balona et al. (2011) did not find any star lying between red end of the SPB stars and blue end of $\delta$ Scuti type stars.

On the basis of distribution of variables in the $H-R$ diagram (Fig. 10) we classified 3 stars as $\beta$ Cep, 25 stars as SPB  and 17 stars as new class. The classification is
mentioned in the last coloumn of Table 3.
Present sample of MS variable stars consists of 17 stars 
which lie below the red end of SPB variables. These stars might have similar kind of variability characteristics like the new class of variables detected
by Mowlavi et al. (2013) in the case of open cluster NGC 3766. The variables
classified on the basis of of the $H-R$ diagram (cf. Fig. 10) are given
in the last column of Table 3.
The periods of new class variables range from 0.17 to 0.58 d. The amplitude of their brightness variation varies from 0.007 to 0.019 mag.
Out of 17 stars four stars numbered V74, V91, V104 and 
V109 were detected spectroscopically as B-type star by Marco \& Negueruela (2002 a, b). 
The detection of 17 new class variables supports the finding of Mowlavi et al. (2013). 

Twenty five MS stars are lying in the instability region of 
SPB stars. Only 20 stars namely V69, V78, V82, V83, V88, V92, V94, V97, V104, V117, V121, V125, V126, V132, V133, V139, V140, V150, V153 and V156
 classified as SPB stars
have periods in the range of 0.4 to 0.6 d. The amplitude of these stars varies from 0.006 to 0.021 mag.
Their variability characteristics reveal that these could 
SPB type stars. The periods estimated for the remaining MS variable stars (V112, V128, V135, V142 and V157) distributed in the instability region of
SPB stars are less than 0.40 d. 
 Star V133 could be probably classical Be star 
(Marco \& Negueruela 2002b) and it is found to be located where new class of variables has been detected by Mowlavi et al. (2013). This star is located in the centre of the cluster NGC 1893 (see Fig. 1). It has a light-curve similar to the Be type stars.
Be stars are known to be fast rotating stars. These stars occupy the same region of 
B-type stars in the $H-R$ diagram. Therefore it becomes difficult to separate them
from normal B-type stars. 

The present variable sample consists of 3 stars which lie in the 
$\beta$ Cep region in the $H-R$ diagram and these could be most 
probably $\beta$ 
Cep type variables.
Majewska-$\acute{S}$wierzbinowicz et al. (2008) presented photometry of stars in cluster $h$ and $\chi$ Persei (NGC 869 and 884) 
 and found a number of $\beta$  Cep,  B and Be type stars. 
They found some stars which show multiperiodic pulsations, indicating $g$ modes or other modes that occur in the
presence of fast rotation. 
\setcounter{figure}{8}
\begin{figure}
\includegraphics[width=9cm]{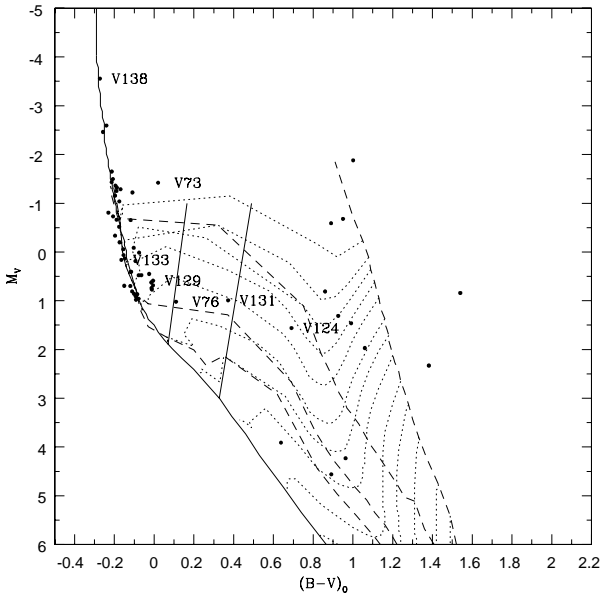}
\caption{ $M_{V}/(B-V)_{0}$ colour-magnitude diagram for the MS and PMS stars of the cluster NGC 1893. The ZAMS by Girardi et al. (2002)
 and PMS isochrones for
0.1, 1, 5, 10 Myrs by Siess et al. (2000) are shown.
The dotted curves show PMS evolutionary tracks of stars of different masses. The classical instability strip of Cepheids is taken from the literature
(see Zwintz \& Weiss 2006). The stars labeled are discussed in section 6.} 
\end{figure}
The cause of variability in B type stars can be understood in terms of Opacity ($\kappa$) mechanism which is  
 a process that drives pulsations in stars.  
Opacity mechanism has been discussed by Eddington (1917), Zhevakin (1953), Cox (1958) and is related to the opacity behaviour in ionisation zones.
Ionisation regions or layers
correspond to a strong increase in opacity which leads the opacity bumps (see Gastine \&  Dintrans 2008). 
The radiation coming from the deeper layers gets blocked in the high $\kappa$ region. The gas heats and the pressure rises below the
layer, pushing it outwards. The
layer expands, and
becomes more transparent to
radiation. Radiation flows through, gas cools, and pressure
below the layer drops. The layer falls inwards and the
cycle repeats.
In case of B type stars observations and theory suggest that the opacity of Iron (Fe) atoms is responsible to drive the observed pulsation.
In a star Fe bump is an abundance of iron at a depth where the temperature is 
very high (T=2,00,000 K). 
The increase in the opacity of Fe is also known as the Z bump. 
The photometric variability in SPB and $\beta$ Cep variables is caused by $\kappa$ mechanism acting in the metal opacity bump (Dziembowski et al. 1993). Dziembowski et al. (1993) predicted the existence of a large region in the MS band at lower luminosities with the help of opacity mechanism, where high order $g$ modes are unstable with the periods ranging from 0.4 to 3.5 d. They also proposed that in other B-type stars having same period range opacity mechanism remains the same.
Townsend (2005) found that retrograde mixed modes are unstable in mid to late B-type stars, as a results of the same iron-bump opacity mechanism
that is usually associated with SPB and $\beta$ Cep stars.  
 Le Pennec \& Turck-Chi$\grave{e}$ze (2014) investigate that Iron group opacities in the envelope of massive stars seems puzzling.    
The possible causes responsible for periodic brightness variation in case of most of the MS stars detected in the present study  
could be pulsations, which occur due to either of fast rotation that somehow alters the internal structure of the star or it could be
result of iron-bump opacity mechanism.
However, more continuous photometry and spectroscopic observations of NGC 1893 are needed to arrive at any conclusion.

Star V138 is the brightest star in our sample.
The spectroscopic study by Marco \& Negueruela  (2002a) assigned this star a spectral type of
O7 V(f).
Whereas its absolute magnitude $M_{V}$ comes out to be
-3.57 mag which results effective temperature of this star as log $T_{eff}$=4.486 to 4.405
(Schmidt-Kaler 1982). Thus, derived effective temperature suggests that it might be type B0 or
B1. The effective temperature of this star using relation between intrinsic $(B-V)_{0}$ colour and effective temperature by Torres (2010)
 comes out to be log $T_{eff}$= 4.428.
 It shows $\beta$ Cep-like pulsation.
MS stars of spectral type O9 to B2 are very interesting targets to study the internal stellar structure by interpreting the observed oscillation characteristics. These are believed to be progenitors of core-collapse supernovae and chemically enrich the Universe (Saesen
et al. 2013).
A number of O-type $\beta$ Cep pulsators have also been detected
in ground-based observations (e.g. Telting et al. 2006; De Cat
et al. 2007; Pigulski \& Pojma$\acute{n}$ski 2008).
Blomme et al. (2011) pointed out that the behaviour of the early-type O stars (spectral types O4-O8) is considerably different from that of later-type stars.
They found that the earliest O stars show red
noise, while the later O-types have pulsational frequencies of the
$\beta$ Cep type. The switch-over occurs around spectral type O8.
The specific physical cause of this noise in O-type stars is
not clear at present. Previous studies found pulsations in late O-type star while
in early O-type star brightness variation could be most possibly because of red noise component.
 Degroote
et al. (2010) detected pulsations in HD 46149 (spectral type O8) and
showed that the observed frequency range and spacings are compatible with theoretical predictions.
HD 46202
(O9 V) shows clear $\beta$ Cep-like pulsations (Briquet et al. 2011).
HD 47129
(O8 III/I + O7.5 V/III) with a series of frequencies
might be a pulsator. But this star has presence of a clear red-noise component
as well (Mahy et al. 2011).
Observations with the MOST 
 satellite of the O9.5 V star $\zeta$ Oph also show $\beta$ Cep-type
pulsations (Walker et al. 2005).
The shape of light curves, amplitude and period suggest that star V138 could be a $\beta$ Cep type star.
In the $H-R$ diagram it is found to be located near the instability strip of the $\beta$ Cep variables.

\setcounter{figure}{9}
\begin{figure}
\includegraphics[width=9cm]{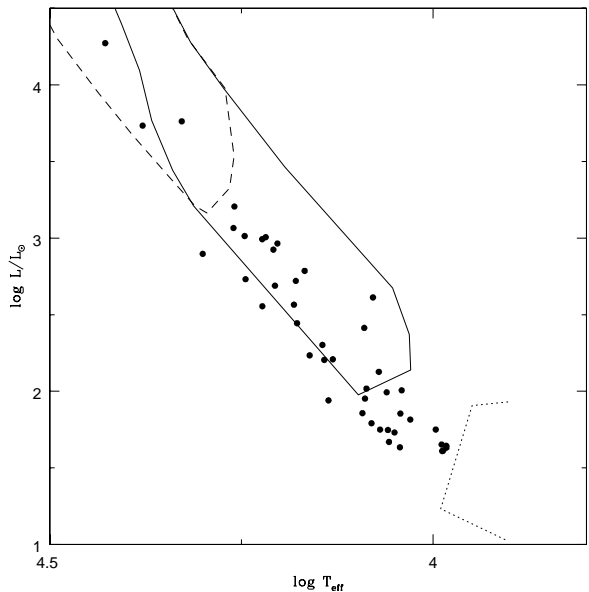}
\caption{ $\log(L/L_{\odot})/ \log T_{eff}$ diagram for the cluster NGC 1893. The theoretical SPB instability strip (continuous curve), 
 empirical $\delta$ Scuti instability strip (dotted curve) and the location of $\beta$ Cep stars (dashed curve) are taken from the literature (see Balona et al. 2011; references therein). 
}
\end{figure}

\subsection{PMS variables}
The majority of the PMS variables detected in the present study have masses $\lesssim$ 3 M$_{\odot}$ and these variables could be TTSs. The estimated periods of these probable TTSs are in the range of 0.187 to 0.528 d. 
The amplitude of the probable PMS variable stars ranges from 0.007 and 0.079 mag. 
The periodic variations in TTSs
are believed to occur due to the axial rotation
of a star with an inhomogeneous surface, having either hot or cool spots (Herbst et al. 1987, 1994).
These stars follow the relation of amplitude and mass as mentioned in Lata et al. (2011, 2012).

Stars V129 and V131 were identified as PMS stars by  Marco \& Negueruela (2002b).
Star V129 could be B-type MS star on the basis of its location in $(J-H)/(H-K)$ TCD.
Its variability characteristics also reveal it to be a B-type pulsating star.
The location of star V131 in $(J-H)/(H-K)$ TCD also suggests it to be a PMS star.
It could be an Herbig Ae/Be star. Its light curve shows periodic variation of about 0.208 d.
Star V73 could also be an Herbig Ae/Be star
on the basis of its location in $(J-H)/(H-K)$ TCD and variability characteristics.
The location of star V76 in $(J-H)/(H-K)$ TCD also suggests that it could also be an Herbig Ae/Be star.
In $M_{V}/(B-V)_{0}$ colour magnitude diagram V76 and V131 lie in the classical instability strip of Cepheids.
It is believed that the Herbig Ae/Be stars show photometric variability
as they move across the instability region in the $H-R$ diagram on their way to the MS (see Lata et al. 2011 and references therein).
Star 124  was selected as the emission-line PMS object because of the presence of emission lines in its spectrum by  Marco \& Negueruela (2002b).
They also find that its spectrum shows weak H$\alpha $ in emission and its spectral type ranges F9V or G0V.
In the $(J-H)/(H-K)$ TCD it is found to be lying in the `F' region where field stars or Class III sources are found to be located.

\subsection{Field population: variable stars}
The present variable sample consists of 43 variable candidates which could be field star population towards the direction of NGC 1893 region.
The $(U-B)/(B-V)$ TCD for the field stars shown in Fig. 11 indicates that these variables are foreground population with $E(B-V)\sim$0.30 mag.
The variability characteristics of field star population indicates that all of them are short
period variables with their amplitude ranging from 0.01 to 0.005 mag. 
These could be short period variables like $\delta$ Scuti stars. 

\setcounter{figure}{10}
\begin{figure}
\includegraphics[width=9cm]{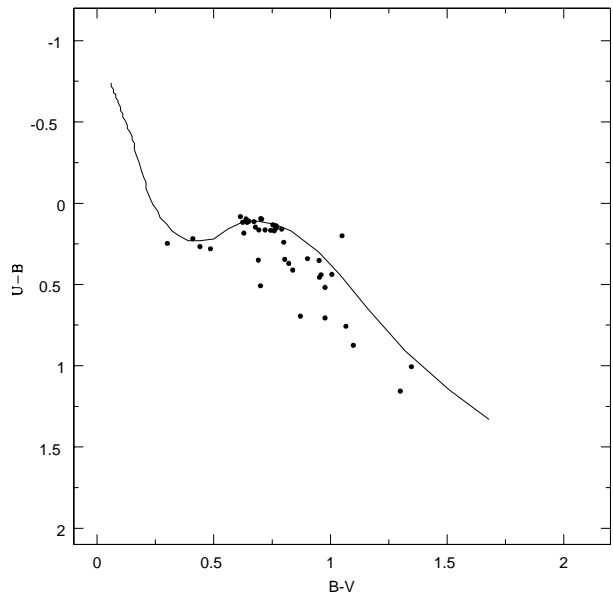}
\caption{The $(U-B)/(B-V)$ TCD for the field stars. The continuous curve represents ZAMS by Girardi et al. (2002) for the foreground field population having $E(B-V)$= 0.30 mag.}
\end{figure}

\setcounter{figure}{11}
\begin{figure}
\includegraphics[width=9cm]{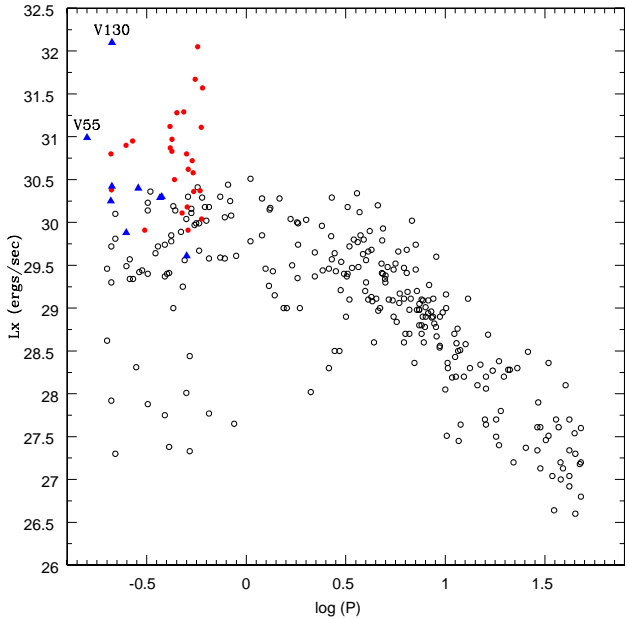}
\caption{ X-ray luminosity vs period for variable stars. Filled circles (MS stars), filled triangles (field stars) and open circles represent present sample and that taken from Pizzolato et al. (2003).}
\end{figure}

\section {X-ray Luminosity vs. Period}
Most of the MS stars are known to emit X-rays. It is believed that massive and hot stars emit X-rays due to shocks 
in their winds or collisions between the wind and circumstellar material (see Wright et al. 2011). In late type 
MS stars the stellar magnetic dynamo is thought to be responsible for the X-rays. A strong correlation between X-ray 
luminosity and rotation period of the late type MS stars has been found (Walter \& Bowyer 1981; Pallavicini et al. 1981; Pizzolato et al. 2003).  
Though the present MS variable candidates belong to upper region of the MS in the $H-R$ diagram,  
 we explored the correlation between X-ray luminosity ($L_{X}$) and period of MS stars (filled circles) and field stars (filled triangles) in Fig. 12. The $L_{X}$ values have been taken 
from the Caramazza et al. (2012). 
In Fig. 12 we have also plotted the $L_{X}$ and rotation period for stars (open
circles) taken from Pizzolato et al. (2003). A comparison indicates
that present data seem to follow the general trend represented by the Pizzolato et al. (2003), however the $L_{X}$ of the present sample is systematically
higher as compared to that of Pizzolato et al. (2003) sample. It is
expected as the star in the present sample has mass 2.7$\lesssim M/M_{\odot}\lesssim$16.0 whereas the sample by Pizzolato et al. (2003) has mass in the range 
of 0.6 to 1.3 $M_{\odot}$. It is well known that $L_{X}$ depends on the
stellar mass (see e.g. Pizzolato et al. 2003; Pandey et al. 2014). The X-ray emission of present sample lies in the saturated region, hence it appears to be dependent on the characteristics of
the stellar structure (cf. Pizzolato et al. 2003). Fig. 12 also indicates that a few field stars like V130 and V55 have higher $L_{X}$, suggesting that these field stars could be massive ones and these could be most probably B type variables.  

\section{Summary}
The paper presents the light curves of 104 variable candidates in the
young open cluster NGC 1893. Among 104 variable candidates 45 stars
could be
MS variables. 
 The periods of MS variables ranges from $\sim$0.15 to $\sim$0.6 d with brightness variation of $\lesssim$ 0.02 mag. 
We classified 3 stars as $\beta$ Cep, 25 stars as SPB stars and 17 stars as new class type stars (cf. Mowlavi et al. 2013) on the basis of their location in the $H-R$ diagram.
We have also found 16 stars as variables which could be of PMS nature. 
Additionally, there are 43 stars which might belong to the field star population.
The correlation between X-ray luminosity ($L_{X}$) and period of MS stars
follows the general trend as given by Pizzolato et al. (2003).

\section{Acknowledgment}
We are thankful to the anonymous referee for useful comments which have improved the contents and presentation of the paper. 

\bibliographystyle{mn2e}

\end{document}